\documentstyle[12pt,epsf]{article}
\newcommand{\s}[1]{\section{#1}\renewcommand{\theequation}
        {\mbox{\arabic{section}.\arabic{equation}}}\setcounter{equation}{0}}

\newcommand{\app}[1]{\section{#1}\renewcommand{\theequation}
        {\mbox{\Alph{section}.\arabic{equation}}}\setcounter{equation}{0}}
\renewcommand{\date}[1]{\par\bigskip\par\sl\hfill #1\par\medskip\par\rm}
\newcommand{\pacs}[1]{\smallskip\noindent{\sl PACS number(s):
                       \hspace{0.3cm}#1}\par\bigskip\rm}
\def\babs{\hrule\par\begin{description}\item{Abstract: }\it} 
\def\eabs{\par\end{description}\hrule\par\medskip\rm}
\newcommand{\ack}[1]{\par\section*{Acknowledgments} #1} 
\renewcommand{\thanks}[1]{\footnote{#1}}

\renewcommand{\title}[1]{\begin{center}\Large\bf #1 \end{center}\rm\par\bigskip}
\renewcommand{\author}[1]{\begin{center}\Large #1\end{center}}
\newcommand{\address}[1]{\begin{center}\large #1 \end{center}}

\begin{document}
\newcommand{\reals}{\mbox{${\rm I\!\!R }$}}
\newcommand{\nats}{\mbox{${\rm I\!\!N }$}}

\newcommand{\cab}{{\cal B}}
\newcommand{\can}{{\cal N}}
\newcommand{\cam}{{\cal M}}
\newcommand{\caz}{{\cal Z}}
\newcommand{\cao}{{\cal O}}
\newcommand{\cac}{{\cal C}}
\newcommand{\cah}{{\cal H}}

\newcommand{\al}{\alpha}
\newcommand{\be}{\beta}
\newcommand{\de}{\delta}
\newcommand{\ep}{\epsilon}
\newcommand{\ga}{\gamma}
\newcommand{\la}{\lambda}
\newcommand{\om}{\omega}
\newcommand{\ze}{\zeta}

\newcommand{\De}{\Delta}
\newcommand{\Ga}{\Gamma}
\newcommand{\Om}{\Omega}
\newcommand{\Si}{\Sigma}

\newcommand{\rS}{{\rm S}}

\newcommand{\snnu}{\sum_{n=0}^{\infty}}
\newcommand{\slnu}{\sum_{l=0}^{\infty}}
\newcommand{\snuu}{\sum_{n=-\infty}^{\infty}}
\newcommand{\sluu}{\sum_{l=-\infty}^{\infty}}
\newcommand{\sied}{\sum_{i=1}^3}
\newcommand{\sneu}{\sum_{n=1}^{\infty}}
\newcommand{\sleu}{\sum_{l=1}^{\infty}}
\newcommand{\sieik}{\sum_{i_1,...,i_k=0}^{\infty}}

\newcommand{\pxi}{\prod_{i=1}^3\left(1-e^{-x_i n}\right)}
\newcommand{\pxj}{\prod_{j=1}^3\left(1-e^{-x_ j n }\right)}

\newcommand{\liar}{\left(e^{-\ep\al}\right)}
\newcommand{\en}{e^{-n\ep\al}}
\newcommand{\enx}{e^{-n\ep\al-nx_i}}
\newcommand{\exi}{(1-e^{-nx_i})}
\newcommand{\exj}{(1-e^{-nx_j})}
\newcommand{\xij}{x_1x_2+x_1x_3+x_2x_3}
\newcommand{\xp}{x_1x_2x_3}
\newcommand{\zrz}{\zeta_R (2)}
\newcommand{\zrd}{\zeta_R (3)}
\newcommand{\zrv}{\zeta_R (4)}
\newcommand{\cont}{\int\limits_{c-i\infty}^{c+i\infty}d\alpha\,\,}

\newcommand{\ent}{d (\nu )}
\newcommand{\pnenj}{\prod_{j=1}^p\nu_j}
\newcommand{\nn}{\nonumber}
\renewcommand{\theequation}{\mbox{\arabic{section}.\arabic{equation}}}
\newcommand{\intsi}{\int\limits_{\Sigma}d\sigma_x\,\,}
\newcommand{\back}{\bar{\Phi}}
\newcommand{\coba}{\bar{\Phi}^{\dagger}}
\newcommand{\abl}{\partial}
\newcommand{\pa}{\partial}
\newcommand{\qpi}{(4\pi)^{\frac{q+1} 2}}
\newcommand{\snenp}{\sum_{n_1,...,n_p=0}^{\infty}}
\newcommand{\tint}{\int\limits_0^{\infty}dt\,\,}
\def\beq{\begin{eqnarray}}
\def\eeq{\end{eqnarray}}
\newcommand{\zb}{\zeta_{{\cal B}}(}
\newcommand{\rzb}{Res\,\,\zb}
\newcommand{\fr}{\frac}
\newcommand{\sip}{\frac{\sin (\pi s)}{\pi}}
\newcommand{\rzs}{R^{2s}}
\newcommand{\g}{\Gamma\left(}
\newcommand{\ikma}{\int\limits_{\ga}\frac{dk}{2\pi i}k^{-2s}\frac{\pa}{\pa k}}
\newcommand{\suani}{\sum_{a=0}^i}
\newcommand{\zem}{\zeta_{{\cal M}}}
\newcommand{\hem}{A^{\cam}}
\newcommand{\hen}{A^{\can}}
\newcommand{\man}{{\cal M}}
\newcommand{\pold}{D^{(d-1)}}
\newcommand{\zesd}{\zeta_{S^d}}
\newcommand{\fac}{\frac{(4\pi)^{D/2}}{a^d|S^d|}}
\newcommand{\sri}{\sum_{i=1}^ d  r_i}
\newcommand{\pri}{\prod_{i=1}^d r_i}
\newcommand{\ber}{B^{(d)}}
\newcommand{\ar}{a|\vec r )}


\title{Bose-Einstein condensation of atomic gases in a
general harmonic oscillator
confining potential trap}
\author{
Klaus Kirsten\cite{kk}
}
\address{Universit{\"a}t Leipzig, Institut f{\"u}r Theoretische Physik,\\
Augustusplatz 10, 04109 Leipzig, Germany}
\author{David J.~Toms\cite{djt}
}
\address{Department of Physics,\\
The University of Newcastle Upon Tyne, Newcastle Upon Tyne, NE1 7RU, England}

\date{March 1996, revised May 1996}
\babs
We present an analysis of Bose-Einstein condensation for a system of non-interacting spin-0 particles in a harmonic oscillator confining potential trap. We discuss why a confined system of particles differs both qualitatively and quantitatively from an identical system which is not confined. One crucial difference is that a confined system is not characterized by a critical temperature in the same way as an unconfined system such as the free boson gas. We present the results of both a numerical and analytic analysis of the problem of Bose-Einstein condensation in a general anisotropic harmonic oscillator confining potential. 
\eabs
\s{Introduction.}
One of the most interesting properties of a system of bosons is that under
certain conditions it is possible to have a phase transition at a critical
value of the temperature in which all of the bosons can condense into the
ground state. It is now well over seventy years since the phenomenon
referred to as Bose-Einstein condensation (BEC) was first predicted 
for the ideal nonrelativistic Bose gas
\cite{Bose,Einstein}. Nowadays it is well known to happen if the spatial
dimension $D\geq 3$. (See \cite{landaulifshitz69} for the case $D=3$ and 
\cite{may64} for general $D$.)

Until recently the best experimental evidence that BEC could occur in
a real physical system was liquid helium, as suggested originally by 
London \cite{London38}. However although the behaviour of liquid helium
at low temperatures can be qualitatively described by the free boson gas
model, the detailed behaviour deviates substantially from this simple model.
Physically this is of course because the effects of interactions which are 
neglected in the free boson gas model are important in liquid helium. 
More recently it was suggested \cite{Blatt,Mosk} that BEC could occur
for excitons in certain types of non-metallic crystals (such as CuCl 
for example). There is now good evidence for this in a number of experiments
\cite{Chaseetc}.

An important development in the last year has been the experimental attempts to observe BEC in very cold gases of rubidium \cite{rub}, lithium
\cite{lith}, and sodium \cite{sod}. This experimental work
has stimulated theoretical studies to try to understand the underlying
physics of the situation \cite{BaymPethick,Fetter,DolString}. 
The systems are very dilute and as a first approximation would be expected to be well 
described by a boson gas model with no interactions among the 
atoms. The atoms are confined in complicated
magnetic traps which can be modelled 
by harmonic oscillator potentials. There have been several
studies of BEC in harmonic oscillator confining potentials
\cite{oldone,MITcrowd,PhysLett,GH,HaugRav}. The purpose of our paper is to
examine the condensation of bosons in a harmonic oscillator potential in a
detailed way which does not use the density of states approach of 
Refs.~\cite{MITcrowd,PhysLett,GH,HaugRav}. Unlike the situation for a boson
gas with no external confining potential in free space \cite{HuangPathria}, there does
not exist a critical temperature which signals a phase transition.
However, we will show that there is a temperature at which the specific 
heat has a maximum which can be identified as the temperature at which 
BEC occurs. (A short report of our results was given in Ref.~\cite{KKDJTshort}.)

The fact that a gas of bosons (neglecting interactions) in a harmonic
oscillator potential does not have a phase transition at some critical
temperature is already apparent from the early work of \cite{oldone}.
(This will also be shown below in a very simple way.) A similar situation
occurs for a system of charged bosons in a homogeneous magnetic field; in
three spatial dimensions BEC does not occur in the same way as for the
case where there is no magnetic field \cite{Schafroth}. The same is true
for bosons confined by spatial boundaries \cite{Pathria}. (See also Ref.~\cite{Kac}.) A general
criterion to decide whether or not BEC occurs has been given recently by
us \cite{KKDJT}, and it is easy to show that the criterion is not met for
a system of bosons confined by a harmonic oscillator potential. 

Since the experiments of Refs.~\cite{rub,lith,sod} are claiming to observe BEC, a natural question which arises concerns the exact nature of the
phenomenon. If BEC as found normally for the free boson gas is impossible
for a system of bosons in a confining potential, then in what sense does
BEC occur ? A natural criterion which has been used in systems of finite
size has been to look at the maximum of the specific heat \cite{PathPak}.
We will apply this criterion to the case of bosons confined by a harmonic
oscillator potential. Given the details of the harmonic oscillator 
potential trap, it is possible to calculate a characteristic temperature 
which can be compared with the values found in the experiments.

The paper is organized as follows.
In section 2 we consider a gas of bosons in an isotropic harmonic
oscillator potential. Although the potentials present in the experiment
are anisotropic, we start with this model because it is much easier 
technically and as we will see afterwards, the anisotropy has not much
influence on the critical temperature where the specific heat has its
maximum. Thermodynamical quantities are given in terms of sums resulting
from the grand partition function of the system. In section 3 we present the 
technique of how to obtain approximate analytic results for all sums involved.
These techniques are used in section 4 to derive simple expressions for the
thermodynamical quantities which allow for a determination of the
critical temperature and the groundstate occupation number. Combining our
analytical techniques with numerical calculations, we present the 
detailed behaviour of the specific heat and other quantities of interest. 
In section 5 we generalize our model to the example of an anisotropic 
oscillator. Once more the detailed behaviour of several quantities relevant
for the recent experiments is given. Appendices A-D contain several technical
details on how to obtain the approximate results for all quantities of
interest. Finally, the conclusions summarize all important results, give a brief comparison between our method and that of Refs.~\cite{PhysLett,GH}, and present a short discussion of further work.

\s{The isotropic harmonic oscillator potential}
For mathematical 
simplicity let us start with the case of an isotropic harmonic oscillator
potential. 
We will assume that the system can be 
described by a grand canonical ensemble. The grand potential is defined by
\beq
q=-\sum_N\ln\left(1-z\exp(-\beta E_N)\right)\;,\label{eq1}
\eeq
where $\beta=(kT)^{-1}$, $E_N$ are the energy levels, and $z=e^{\beta\mu}$ 
is the fugacity in terms of the chemical potential $\mu$. It proves 
convenient to expand the logarithm in (\ref{eq1}) to obtain
\beq
q=\sum_{n=1}^{\infty}\frac{z^n}{n}\sum_N\exp(-n\beta E_N)\;.\label{eq2}
\eeq
For an isotropic harmonic oscillator characterized by an angular frequency 
$\omega$ the energy levels are given by 
$E_{n_1n_2n_3}=\hbar\om (n_1+n_2+n_3+3/2)$, $n_1,n_2,n_3\in\nats_0$. If we set $n_1+n_2+n_3=k$, where $k\in\nats_0$, then the
energy levels may be ordered in the way
$E_k=(k+3/2)\hbar\omega$ with 
multiplicity $(k+1)(k+2)/2$. The sum over $N$ 
in (\ref{eq2}) may be performed to obtain
\beq
q=\sum_{n=1}^{\infty}\frac{e^{n\beta(\mu-3/2\hbar\omega)}}{n(1-e^{-nx})^{3}}
\;,\label{eq3}
\eeq
where we have defined the dimensionless variable $x=\hbar\omega/(kT)$. 
The number of particles is given by 
$\displaystyle{N=\beta^{-1}\left(
\frac{\partial q}{\partial\mu}\right)_{T,\omega}}$, which becomes
\beq
N=\sum_{n=1}^{\infty}\frac{e^{n\beta(\mu-3/2\hbar\omega)}}{(1-e^{-nx})^3}
\;,\label{eq4}
\eeq
when (\ref{eq3}) is used.

In order that the number of particles remains positive, it is necessary for
$\mu\le (3/2)\hbar\omega$. (More generally, we require $\mu\le E_0$ where
$E_0$ is the lowest energy level.) Normally the critical temperature for
BEC is the temperature at which $\mu =E_0$, which for the isotropic
harmonic oscillator reads $\mu=(3/2)\hbar\omega$. It is now easy to see that
BEC cannot occur in the same way for bosons confined in the harmonic
oscillator potential as it does for bosons in free space. In the case of
the free boson gas in free space with no confining potential, as the
temperature is lowered the chemical potential $\mu$ increases from
negative values towards the value 0. (This is in agreement with the
general result $\mu=E_0$ quoted above since the lowest energy level is
zero for the free boson gas.) The value of the temperature at which
$\mu=0$ defines a critical temperature $T_c$ determined in terms of the
particle density. At temperatures lower than $T_c$, $\mu$ remains frozen
at the value $\mu=0$, and the number of particles found in excited states
is bounded. If the total number of particles exceeds this bound then the
only possibility is for the excess particles to be found in the ground
state, giving rise to BEC. This standard scenario is described in
\cite{HuangPathria} in some detail. The phase transition which occurs is related to the breaking of the $U(1)$ gauge symmetry associated with the change of phase of the Schr\"{o}dinger field. We have discussed this in a recent review \cite{KKDJTreview}.

The origin of the different behaviour between the confined and the free Bose gas might be seen more clearly by
considering the number of particles in the ground state with energy
$(3/2)\hbar\om$.
In addition to the dimensionless quantity $x$ we introduce $\mu =\hbar \om
(3/2 -\ep)$, the limit $\ep \to 0$ corresponding to the limit of the 
chemical potential reaching its critical value. 
In terms of $x$ and $\epsilon$, the number of particles in the groundstate is
\beq
N_{ground}=\frac 1 {e^{\ep x}-1}.\label{ground}
\eeq
For $\ep\to 0$ we have $N_{ground}\to\infty$, this being 
essentially the reason that no BEC in the usual sense that $\ep$ reaches
$0$ at some finite temperature might occur. For given $x$ and particle 
number $N$ it is clear from (\ref{ground}) that $\ep>(1/x)\ln((N+1)/N)$ 
and that $\ep$ can reach $0$ only in the zero temperature limit or in the limit
$N\to\infty$. However, as is also
clear from (\ref{ground}), for fixed particle number $N$, once $\ep$ is small
enough, it is essentially only the groundstate which is occupied. Lowering
the temperature, which is the same as increasing $x$, this will happen unavoidably as can be 
seen from (\ref{eq4}). The temperature at which the groundstate starts to
increase considerably its occupation number is the most dramatic moment 
in the system. The extreme behaviour is similar but not equal to a phase 
transition. Although quantities are changing rapidly, everything behaves
smoothly; no discontinuities appear.  
The argument presented here may be performed in a much more general context 
using the setting described in \cite{KKDJT}. Further discussion of this important point we postpone until we have presented our analytical analysis of the thermodynamical quantities.

Let us continue with the internal energy $U$ of the system which is given in terms of the grand potential $q$ by
\beq
U=\left\{-\frac{\pa}{\pa \be} +\frac{\mu}{\be} \frac{\pa}{\pa \mu}
\right\} q. \label{eq5}
\eeq
In terms of the dimensionless quantities $x$ and $\epsilon$,
$U$ reads
\beq
\frac U {\hbar\omega}=-\frac{\partial q}{\partial x}-
\frac{(3/2-\epsilon)}{x}\frac{\partial q}{\partial\epsilon}\;.
\label{eq6}
\eeq
Using the series for $q$ given in (\ref{eq3}) results in
\beq
\frac U {\hbar\omega}=\frac{3}{2}N+3u_1\;,\label{eq7}
\eeq
where 
\beq
u_1=\sum_{n=1}^{\infty}e^{-n\epsilon x -nx}(1-e^{-nx})^{-4}\;.
\label{eq8}
\eeq
In terms of $U$, the specific heat reads
\beq
C=\left(\frac{\partial U}{\partial T}\right)_{N,\omega
\mbox{ held fixed}}\;.\label{eq9}
\eeq
Since $N$ is held fixed when computing $C$, 
only $u_1$ contributes to the specific heat, and we find 
\beq
C/k=-3x^2\left(\frac{\partial u_1}{\partial x}
\right)_{N,\omega\mbox{ held fixed}}\;.\label{eq10}
\eeq
Once more due to fixed $N$, alternatively one might use
\beq
\tilde u _1 = \sneu \frac{e^{-\ep n x}}{\left(1-e^{-nx}\right)^4}
,\label{eq11}
\eeq
which differs from $u_1$ only by a multiple of $N$. Continuing with
(\ref{eq10}) one first finds
\beq
\left(\frac{\pa u_1}{\pa x}\right) _
{N,\om \mbox{ held fixed}} =-S_2 \left[ 1+\frac{\pa}{\pa x} 
(\ep x )_{N,\om}\right]-4S_3,\label{eq12}
\eeq
with
\beq
S_2&=&\sum_{n=1}^{\infty}ne^{-n\epsilon x-nx}(1-e^{-nx})^{-4}
\;,\label{eq13}
\eeq
and
\beq
S_3&=&\sum_{n=1}^{\infty}ne^{-n\epsilon x -2nx}(1-e^{-nx})^{-5}
.\label{eq14}
\eeq
Differentiating equation (\ref{eq4}) with respect to $x$ for fixed
$N,\om$, $\displaystyle{\frac{\pa (\ep x)}{\pa x} }$ is determined. We find
\beq
\left(\frac{\partial(\epsilon x)}{\partial x}
\right)_{N,\omega\mbox{ held fixed}}=-3\frac{S_2}{S_1}\;,
\label{eq15}
\eeq
where in addition to $S_2$ and $S_3$ we introduced
\begin{eqnarray}
S_1&=&\sum_{n=1}^{\infty}ne^{-n\epsilon x}(1-e^{-nx})^{-3}.\\
\end{eqnarray}
So putting the results of equations (\ref{eq10}), (\ref{eq12}) and 
(\ref{eq15}) together, we arrive at
\beq
C/k=3x^2\left\lbrace 4S_3+S_2-3\frac{S^2_2}{S_1}\right\rbrace\;.
\label{eq17}
\eeq

Before we proceed with the numerical analysis of some of the above
thermodynamical quantities, we turn now to the analytical treatment
of 
these quantities for some range of parameters $x$ and $ \ep$.
Let us look at the relevant range of parameters in the sodium
experiment. (Qualitatively it will be the same for the other experiments.) 
There \cite{sod} we have $\om/(2\pi) =416$ Hz if we use the geometric mean of
the frequencies. The relevant temperature range is around $2\mu K$. It is therefore 
seen, that the behaviour of the thermodynamical quantities for small $x$ is desired. A plausible approach is
to argue that for $x\ll 1$, it is justified to replace sums which have arisen in the expansions above with integrals. Care must be exercised with this to take a proper account of the density of states. (We will return to this in Sec.~6.) This is tantamount to regarding the energy levels as continuous rather than
discrete. However, we have shown recently that the behaviour of 
thermodynamical systems with a discrete energy spectrum is completely different from one with a continuous energy spectrum. In the first case, no real BEC can occur whereas in the
second case it does \cite{KKDJT}. (By real BEC, we mean that there is a phase transition such as that which occurs in the free boson gas.)
If the correct behaviour for small $x$ is desired, 
one approach which is definitely safe is to deal with the exact sums. The sums do not converge
very rapidly for small $x$, nor do they display in any transparent way the 
behaviour at small $x$. However, it is possible to convert the sums into  
contour integrals, and by deforming the contours of integration in an
appropriate way obtain at least asymptotic expansions for some
appropriate range of the parameters. The details of this procedure will be described in the following section.

\s{Analytical treatment of harmonic oscillator sums}
Let us now describe in some detail the analytical treatment of the
sums appearing in the thermodynamical quantities. As one can see, all
sums involved are of the form
\beq
f(l,k,m) = \sneu \frac{e^{-n\ep x-mnx}}{n^l \left(1-e^{-nx}\right)^k}
,\label{eq18}
\eeq
with different integral values for $l,k,m$. Let us consider the range of $x\ll 1
$,
which is actually fulfilled in the recent experiments described below,
and in addition $\ep \ll 1$ corresponding to the range where a phase 
transition occurs in $3$ dimensional free space and, as we will see, 
corresponding to the range very close to the maximum of the specific heat.

Probably the best technique for the analysis of (\ref{eq18}) in the 
mentioned range of parameters is the use of the Mellin-Barnes integral
representation. We are going to apply it in its simplest form, making use
of
\beq
e^{-v} = \frac 1 {2\pi i}\int\limits_{c-i\infty}^{c+i\infty} d\alpha
\,\,\Gamma (\al ) v^{-\al},\label{eq19}
\eeq
valid for $\Re v > 0$ and $c\in \reals$, $c>0$. Equation (\ref{eq19}) is 
easily proven by closing the contour to the left obtaining immediately 
the power series expansion of $\exp (-v)$. In order to apply equation
(\ref{eq19}) write (\ref{eq18}) in the form
\beq
f(l,k,m)& =& \sneu n^{-l} \sieik e^{-nx\left[ \ep+m+i_1+...+i_k\right]}\nn\\
&=&\sneu n^{-l} \sieik \frac 1 {2\pi i} 
\cont \Gamma (\al) n^{-\al} x^{-\al} \left[\ep+m+i_1+...+i_k\right]^{-\al}
.\label{eq20}
\eeq
Now we would like to interchange the summation and integration in order
to arrive at an expression in terms of known zeta functions, in detail
the Riemann zeta function $\zeta_R (s)$,
\beq
\zeta_R (s) =\sneu n^{-s},\label{eq21}
\eeq
and a Barnes zeta function \cite{barnes03},
\beq
\zb s,a,k) = \sieik \left[ i_1+...+i_k +a\right] ^{-s}.\label{eq22}
\eeq
The basic properties of the Barnes zeta function are summarized in the 
Appendix A. In order to allow for the interchange of summations and 
integration one has to ensure the absolute convergence of the resulting
sums \cite{weldon86,elizalderomeo89}. This is certainly true for $\Re c > {\rm max}\,\,(k,1-l)$ and one 
arrives at
\beq
f(l,k,m) =\frac 1 {2\pi i} \cont \Gamma (\al ) x^{-\al} \zeta_R (\al +l)
\zb 
\al , m+\ep,k)
.\label{eq23}
\eeq
This is a very suitable starting point for the analysis of certain properties
of the sums $f(l,k,m)$. Closing the contour to the right corresponds to the
large-$x$ expansion; closing it to the left to the small-$x$ expansion. To the
right of the contour the integrand  in (\ref{eq23}) has no poles, which means that the large-$x$ 
behaviour contains no inverse power in $x$. One might show however, that the
contribution from the contour itself is not vanishing at infinity leading to exponentially damped
contributions for $x\to\infty$, the well known behaviour of partition sums 
at low temperature.

As mentioned, this is not the range of interest for recent experiments and
we concentrate on the small-$x$ behaviour thus closing the contour to the 
left. Closing to the left there appear to be three different 
sources of poles,\\
i.) poles of $\zb \al , m+\ep ,k)$ for $\al =1,...,k$; (See Appendix A.)\\
ii.) pole of $\zeta_R (\al +l)$ for $\al =1-l$;\\
iii.) poles of $\Gamma (\al )$ for $\al =-p$, $p\in\nats_0$.
\\
Depending on the value of $l$ there might be a double pole at $\al =1-l$. In detail for 
$l=0,-1,...,1-k$, there is a double pole from i.) and ii.); for $l=1,...,
\infty$ there is a double pole from ii.) and iii.). Collecting all poles 
is an easy exercise for all values of $l$. We will restrict to the relevant
values of $l$ for our problem, these being $l=1,0,-1.$ 

For $l=1$ we find
\beq
f(1,k,m) &=& \sum_{n=1}^k \Gamma (n) x^{-n} \zeta_R (1+n) \rzb n, \ep +m,k)
\label{eq24}\\
& & +\zeta_{\cab}' (0,m+\ep,k)-(\ln x) \zb 0,\ep+m,k)+\cao (x),\nn
\eeq
$\mbox{Res}\,\,\zeta_{\cab}$ being 
the residue of the Barnes zeta function, and $\zeta_{\cab}'$
its derivative with respect to $s$. (See equation (\ref{eq22}).) The residues
and values of the Barnes zeta function are derived in Appendix~A.
The leading important residues are
\beq
\rzb k,a,k) &=&=\frac 1 {(k-1)!},\nn\\
\rzb k-1,a,k) &=&\frac{k-2a}{2(k-2)!},\label{resbarnes}\\
\rzb k-2,a,k) &=& \frac{6a^2 -6ak+(k/2) (3k-1)}{12 (k-3)!}.\nn
\eeq
Also the derivative of $\rzb s,a,k)$ with respect to $s$ at $s=0$ might be determined in terms of derivatives of
the Hurwitz zeta function, but the contributions are of subleading order
and will not be used here and thus are not presented.

For $l=0,-1$, the formula analogous to (\ref{eq24}) reads
\beq
f(l,k,m) &=& \sum_{n=1\atop n\neq 1-l}^k \Gamma (n) x^{-n} \zeta_R (l+n)
\rzb n,m+\ep,k) \nn\\
& &+x^{l-1}\Gamma (1-l) \left\{ \mbox{PP}\,\,\zb 1-l,m+\ep,k) \right.\nn\\
& &\left. \qquad\qquad+(\gamma -\ln x +\psi (1-l))
 \rzb 1-l,m+\ep,k) \right\}\nn\\
& &+\zeta_R (l) \zb 0,m+\ep,k) +\cao (x),\label{eq25}
\eeq
PP $\zeta_{\cab}$ 
denoting the finite part of $\zeta_{\cab}$, $\psi (x) =(d/dx) \ln \Gamma (x)$
and $\psi (1) =-\gamma$. 

The presented asymptotics together with (\ref{a7}) allow one to obtain the
small $x$ and $\ep$ behaviour of the theory. In the next section we list the 
asymptotic expansions of the various thermodynamical quantities. The needed sums are given
in the Appendix B for the convenience of the reader.

By changing slightly the procedure described previously, it is also 
possible to obtain an expansion for $x\ll 1$ not restricted to $\ep \ll 1$.
To find this representation write instead of equation (\ref{eq20}) 
\beq
f(l,k,m) =\sneu \frac{e^{-nx\ep }}{n^l} \sieik \frac 1 {2\pi i} 
\cont \Gamma (\al ) n^{-\al}x^{-\al} \left[
m+i_1+...+i_k\right]^{-\al},\label{eqn1}
\eeq
which is found by using equation (\ref{eq19}) only for 
$\exp (-nx[m+i_1+\ldots +i_k ])$ and excluding the part $\exp (-nx\ep )$ from
the procedure. For the case that $m=0$, one has to treat separately the 
zero-mode $i_1=\ldots i_k =0$. Closing the contour once more to the left
to obtain the $x\ll 1$ behaviour, an expansion in terms of the polylogarithm
\beq
Li_n (x) =\sleu \frac{x^l}{l^n}, \label{eqn2}
\eeq
is found. 
The basic properties of the polylogarithm may be found in \cite{19,20}.
For reasons of clarity the corresponding results are 
summarized in Appendix D, however only for the anisotropic harmonic oscillator
presented in section 5. The isotropic case is easily extracted from Appendix D. Because our intention is to try to present simple analytical expressions we will not use the expansion in polylogarithms, since by necessity this would involve numerical evaluation.

\s{Temperature dependence of the thermodynamical quantities}
Having described in detail the application of Mellin-Barnes integral 
techniques for the approximate calculation of harmonic
oscillator sums, we are now prepared to present the results for all
thermodynamical quantities of interest. First of all for the grand potential we have
$q=f(1,3,0)$ and using equation (\ref{eq24}) results in
\beq
q=\frac{\zrv}{x^3} +\frac{\left(\frac 3 2 -\ep\right) \zrd}{x^2}
+\frac{\zrz} x +\cao (\ln x,\ln\ep ).\label{eq26}
 \eeq
For the number of particles we have $N=f(0,3,0)$ and find
\beq
N=\frac{\zrd}{x^3} +\frac{\left(\frac 3 2 -\ep\right)\zrz}{x^2}
+\frac 1 {\ep x}+\cao \left(\frac{\ln x} x \right).\label{eq27}
\eeq
It is also possible to present equation (\ref{eq27}) in a slightly 
different way. As we have explained in some detail in section 2, a quantity
of special interest is the number of particles in the groundstate, 
given by (\ref{ground}). Splitting $N=N_{ground}+N_{excited}$
and using the same techniques as described in section 3, but now with
$i_1=\ldots=i_k=0$ excluded from the summation in equation (\ref{eq20})
(this summation index actually corresponds exactly to the groundstate),
the following asymptotic expansion is found,
\beq
N=N_{ground} +\frac{\zrd}{x^3} +\frac{\left(\frac 3 2 -\ep\right)
\zrz}{x^2} +\cao\left(\frac{\ln x} x\right).\label{groundasymp}
\eeq
Equation (\ref{eq27}) is very useful to determine $\ep$ as a function of
$N$ and $x$. This then leads with the help of equation (\ref{groundasymp})
to the groundstate occupation number as a function of $x$ (for fixed $N$).

Using $u_1 = f(0,4,1)$ and furthermore equation (\ref{eq7}), the asymptotic
expansion of the internal energy reads
\beq
\frac U {\hbar \om} &=& \frac{3\zrv}{x^4} +\frac{\zrd}{x^3}\left(\frac 9 2 
-3\ep \right) +\frac{13\zrz}{4x^2}\nn\\
& &+\frac 3 {2x\ep} +\cao \left( \frac{\ep}{x^2} ,\frac 1 x\right).
\label{eq28}
\eeq
The asymptotics for $u_1$ together with the asymptotics of $S_1,S_2,S_3$ needed
for the analysis of the specific heat are listed in appendix B. After
some calculation we find using (\ref{eq17}) the following result,
\beq
\frac C k &=& \frac{12\zrv}{x^3} +\frac{9\zrd}{x^2} +\frac{2\zrz} x -\frac
{12\ep\zrd}{x^2}\label{eq29}\\
& &-\frac{18\ep^2\zrz \zrd}{x^3 } -\frac{9\ep^2\zrd^2}{x^4} 
+\frac{9\ep^4\zrz\zrd^2}{x^6}+\cao \left(\ln x,\frac{\ep}{x}\right)\;.\nn
\eeq
This concludes the list of the asymptotic behaviour 
for $x\ll 1$, $\ep \ll 1$, of the most important
thermodynamic quantities considered here.

The corresponding expansions in terms of the polylogarithm (see the end
of section 3) are also easily obtained and given in Appendix D. Once more
the results for the isotropic harmonic oscillator follow immediately from those for the anisotropic oscillator.

The above results together with a numerical treatment of the 
thermodynamical quantities using their explicit representations in form
of the sums given in section 2 allows a very detailed prescription over the
whole temperature range of relevance for the recent experiments. As an illustration we will first choose parameters pertinent to the rubidium experiment \cite{rub}. We choose $N=2000$ here. The aim is to compute the chemical potential which is given by $\epsilon$. (Recall that $\mu=\hbar\omega(3/2-\epsilon)$.) This may be done by solving (\ref{eq4}) for $\epsilon$ as a function of $x$. (Recall that $x=\hbar\omega/(kT)$ is the inverse temperature in appropriate dimensionless units.) The numerical result of this calculation is shown as the solid curve in Fig.~1. As can be seen from this figure, $\epsilon$ undergoes a very rapid decrease from values of the order of unity to values of the order of $10^{-2}$ over a very small range of $x$. After this sharp decrease, $\epsilon$ goes asymptotically to zero as $x$ increases ( or as $T$ decreases). This contrasts with a real phase transition such as occurs in the free Bose gas where $\epsilon$ would reach the value $\epsilon=0$ at some nonzero temperature which could be identified with the critical temperature. From (\ref{ground}) it can be seen that this sudden drop in $\epsilon$ is associated with a sudden rise in the ground state occupation number, and is therefore associated with the onset of BEC.

Although the chemical potential has a sudden change, the change happens in a completely smooth way; thus the identification of a specific critical temperature is problematic in this case. One approach which has been used in finite volume systems where similar behaviour occurs \cite{PathPak} is to calculate the maximum of the specific heat and identify the temperature at which the maximum occurs with the critical temperature. In Fig.~2 we illustrate with a solid curve the result of a calculation of the specific heat using the exact harmonic oscillator sums. It is seen to have quite a sharp but smooth maximum at a value $x_m\simeq0.0921$. If we use $\omega/(2\pi)=60$ Hz, as for the strong trap referred to in Ref.~\cite{BaymPethick}, then the specific heat maximum occurs at the temperature $T\simeq3.127\times10^{-8}$ K.

We now turn from a numerical evaluation to the use of our approximate analytical results detailed above. Because our approximation assumed that $x$ and $\epsilon$ were both small we would not expect the results to apply for $x\leq x_m\simeq0.0921$ since Fig.~1 shows that $\epsilon$ is already becoming quite large. If we define $f$ to be the fraction of particles in the ground state,
\beq
N_{ground}=fN\;,\label{new4.1}
\eeq
then from (\ref{ground}) we have
\beq
\epsilon x=\ln\left(1+\frac{1}{fN}\right)\;.\label{new4.2}
\eeq
Using (\ref{groundasymp}) this yields
\beq
(1-f)N\simeq\zeta_R(3)x^{-3}+\big(\frac{3}{2}-\epsilon)\zeta_R(2)x^{-2}
\;.\label{new4.3}
\eeq
$\epsilon$ may be eliminated from (\ref{new4.3}) by using (\ref{new4.2}). This results in a cubic equation which determines $x$ for a given $f$ and $N$. Once $x$ has been found $\epsilon$ is determined by (\ref{new4.2}). We have shown the result of this approximate evaluation of $\epsilon$ as the diamonds in Fig.~1. As expected, once $x$ decreases below $x_m\simeq0.0921$, the agreement between our approximation and the exact result breaks down. However over the region where the specific heat maximum occurs, our approximation for $\epsilon$ is quite good. As $x$ increases (so that the temperature becomes less that the critical temperature), the agreement between our approximate value for $\epsilon$ and the true value becomes better and better. Increasing values of $x$ correspond to an increasing fraction of particles in the ground state. Fig.~1 shows that the agreement between our approximate value for $\epsilon$ and the true value remains remarkably good even up to values of $\epsilon\simeq0.5$. Given that we assumed $\epsilon\ll1$ in our derivations, this is an unexpected result.

Since our approximate result for $\epsilon$ is good in the region where the specific heat maximum occurs, we can have some faith in our other approximate results for $x\ge x_m\simeq0.0921$. In Fig.~2 the diamonds illustrate the result of using our approximate result (\ref{eq29}) for the specific heat. Again the agreement between the approximation and the exact result is seen to be good up to the maximum. For values of $x<x_m$, the approximation breaks down for the reason already mentioned. We have shown a more detailed comparison between our approximation for the specific heat and the true value in Fig.~3. The diamonds illustrate the ratio of our result to the exact value. For $x\approx x_m$ our approximate result is within a few percent of the true value, and for $x>x_m$ the agreement becomes better than 1\%.

We can also compare our results to the bulk results obtained by directly converting the harmonic oscillator sums into integrals as in Refs.~\cite{MITcrowd,HaugRav}. (We will use the terminology bulk rather than thermodynamic limit as in Ref.~\cite{Kac}.) For the specific heat this amounts to just keeping the first term on the right hand side of (\ref{eq29})~:
\beq
C_{bulk}/k=12\zeta_R(4)x^{-3}\;.\label{new4.4}
\eeq
For the particle number the bulk result consists of dropping the term in $x^{-2}$ in (\ref{groundasymp}) along with all subdominant terms, taking
\beq
N_{bulk}=N_{ground}+\zeta_R(3)x^{-3}\;.\label{new4.5}
\eeq
A way of improving the bulk approximation was given in Refs.~\cite{PhysLett,GH}, and we will return to the relationship of this improvement to our approach in Sec.~6.

The bulk transition temperature is obtained by equating $N_{ground}$ to zero. This gives
\beq
x_{bulk}=\left\lbrack\frac{\zeta_R(3)}{N}\right\rbrack^{1/3}\;,
\label{new4.6}
\eeq
where $x_{bulk}=\hbar\omega/(kT_{bulk})$. Eq.~(\ref{new4.4}) holds for $x\ge x_{bulk}$. We have plotted the ratio of $C_{bulk}$ to the exact specific heat as the crosses shown in Fig.~3. Although the agreement between the bulk value and the exact value is quite good near the specific heat maximum, it is off by about 25\%  when $x$ reaches the value 0.4. In contrast our approximation is off by less than 1\%. Another feature of using the bulk approximation is that the specific heat is found to be discontinuous at the bulk temperature, in contrast to the smooth behaviour found in Fig.~2.

A final comparison we will make is for the ground state occupation number. In Fig.~4 the diamonds represent the result of using our approximation for the particle number and the crosses the result of using the bulk value. Both results are shown as a ratio with the exact value found from a numerical evaluation of the harmonic oscillator sum. Close to the specific heat maximum our results are off by about 10\% and the bulk results are off by around 300\%. As $x$ increases, our result converges very rapidly towards the true value, whereas the convergence of the bulk results is slower. For large $x$ both approximations become indistinguishable.

It is possible to obtain an approximate analytic expression for the BEC temperature and compare it to the bulk temperature. Suppose that $x$ is close to the bulk value given in (\ref{new4.6}) and write
\beq
x=x_{bulk}(1+\eta)\;,\label{new4.7}
\eeq
for some small $\eta$. If we assume $\epsilon<<3/2$, then from (\ref{new4.3}) we find
\beq
\eta\simeq\frac{1}{(1-f)}\left\lbrack\frac{1}{3}f+
\frac{\zeta_R(2)}{2\lbrack\zeta_R(3)\rbrack^{2/3}}\,N^{-1/3}
\right\rbrack\;.\label{new4.8}
\eeq
This result assumes that $x_{bulk}$ is small, but makes no assumption about the size of $f$. Typically we find that for particle numbers $N\sim10^3-10^6$, at the point where the specific heat maximum occurs $f$ is a few percent. We can therefore say $(1-f)^{-1}\simeq1$, which leads to
\beq
\frac{T-T_{bulk}}{T}\simeq-
\left\lbrack\frac{1}{3}f+
\frac{\zeta_R(2)}{2\lbrack\zeta_R(3)\rbrack^{2/3}}\,N^{-1/3}
\right\rbrack\;.\label{new4.9}
\eeq
If we put $f=0$ this gives the result in Ref.~\cite{PhysLett,GH}. For $N=2000$ it is easily seen that using the bulk value for the temperature is only about 10\% higher than using the value determined by the maximum of the specific heat. Both temperatures approach one another as the particle number increases. Thus as an estimate of the critical temperature, the use of the bulk value is quite a good approximation. However as our results above show, care must be exercised in using the bulk values for the particle number or specific heat.

To finish our discussion of the isotropic harmonic oscillator we wish to mention briefly some results for other choices of particle number. We would expect that as $N$ increases not only will the bulk approximation become better, but so will ours. We have done the calculations just described for $N=2\times10^4,\ 2\times10^5$ and $5\times10^5$. Because the resulting figures are similar to those already presented in the case $N=2000$ there is little point in showing them here. The expectation of improved agreement between the approximations and the exact result is borne out. The specific heat maximum occurs at $x_m\simeq0.0408$ for $N=2\times10^4$, at $x_m\simeq0.0185$ for $N=2\times10^5$, and at $x_m\simeq0.0136$ for $N=5\times10^5$. These values correspond very closely to those obtained using the approximate formula (\ref{new4.9}).

\s{The anisotropic harmonic oscillator potential}
After having described in detail the isotropic harmonic oscillator 
potential let us explain the new technical problems one encounters when
treating the anisotropic case. 
The calculation parallels very much the one for the isotropic harmonic 
oscillator and we can be brief.
The energy eigenvalues are given by
\beq
E_{n_1n_2n_3} =\hbar \sied \om_i \left( n_i +\frac 1 2 \right),
n_i \in \nats_0 .\label{h1}
\eeq
As we will see in the following, it is useful to introduce the 
dimensionless quantities,
\beq
x_i =\hbar \beta \om_i; \qquad \Om =\frac 1 3 \sied \om_i;\qquad
\al =\hbar \be \Om .\nn
\eeq
We then have
\beq
\be E_{n_1n_2n_3} =\sied n_i x_i +\frac 3 2 \al .\nn
\eeq
In analogy to the harmonic oscillator we use furthermore
\beq
\mu =\hbar \Om \left( \frac 3 2 -\ep \right) \nn
\eeq
to find
\beq
\be E_{n_1n_2n_3} -\be\mu =\sied x_i n_i +\al\ep .\label{h2}
\eeq
In terms of the dimensionless variables the grand potential reads
\beq
q= \sneu \frac{\en}{ n\pxi }. \label{h3}
\eeq
For the particle number we have
\beq
N= \sneu \frac{\en }{ \pxi} .\label{h4}
\eeq
With eqs.~(\ref{h3}) and (\ref{h4}) one easily gets the internal energy,
\beq
\frac{U}{\hbar \Om } =\frac 3 2 N +\sied u_i ,\label{h5}
\eeq
where we defined 
\beq
u_i =\frac{x_i} {\al }\sneu \frac{\enx}{\pxj} \frac 1 {\exi}.\label{h6}
\eeq
Continuing for the specific heat as done for the analysis of the isotropic harmonic 
oscillator, we arrive at
\beq
\frac C k &=& \sied x_i S_{2,i} \left[ x_i -\sum_{j=1}^3 x_j \frac{S_{2,j}}
{S_1} \right]\nn\\
& &+\sied x_i^2 S_{3,ii} \label{h7}\\
& &+\sied \sum_{l=1}^3 x_i x_l S_{3,il},\nn
\eeq
with 
\beq
S_1& =& \sneu \frac{n \en}{\pxi},\nn\\
S_{2,i}& =&\sneu \frac{ n\enx }{\pxj} \frac 1 {\exi} ,\nn
\eeq
and
\beq
S_{3,il} &=& \sneu \frac{n e^{-n\ep\al -n (x_i+x_l)}}{\pxj}
 \frac 1 {\exi} \frac 1 {\left(1-e^{-x_l n}\right)}.\nn
\eeq
The asymptotic expansions of all above quantities may be obtained 
using the same techniques as for
the harmonic oscillator described in section 3. 
The only difference is that one has to deal 
with the slightly more general function 
\beq
\zb s,a|\vec x) =\sum_{\vec m =0}^{\infty} (a+\vec m \vec x )^{-s} .
\label{h8}
\eeq
The asymptotic expansions for the thermodynamical quantities involves the
residues of the function $\zb s,a|\vec r )$, the
basic properties of which are summarized in Appendix A. Using once more the 
Mellin-Barnes integral representation 
in complete analogy to section 3, we arrived at the asymptotics 
for $u_i$, $S_1,$ $S_{2,i}$ and $S_{3,il}$. These are all summarized in
Appendix C. We here list only the asymptotics of the physical quantities,
\beq
q&=& \frac{\zrv}{\xp} +\frac{\zrd \al}{\xp} \left(\frac 3 2 -\ep \right)
\label{h9}\\
& & +\frac{\zrz \al ^2} {\xp} \left( 
\frac{\xij }{12 \al^2} +\frac 3 4 \right) +...,\nn\\
N&=& \frac{\zrd}{\xp} + \frac{\zrz \al}{\xp}\left(\frac{3}{2}-\epsilon\right) +\frac 1 {\ep\al}+...
,\label{h10}\\
\frac{U}{\hbar \Om} &=& \frac{ 3\zrv}{\xp \al} +\frac{3 \zrd}{\xp}
\left(\frac 3 2 -\ep \right) \label{h11}\\
& &+\frac{6 \zrz \al}{\xp}\left( \frac 1 2 +\frac 1 {72} \frac{\xij} {\al^2}
\right)+...\nn\\
\frac C k &=& \frac{12 \zrv}{\xp} +\frac {9\zrd \al}{\xp} 
-\frac{9\al^2 \ep^2 \zrd ^2} {(\xp )^2} \nn\\
& &+\frac 9 4 \frac{\zrz \al^2}{\xp} -\frac 1 {12} \frac{\zrz \vec x ^2}
{\xp} \label{h12}\\
& &-\frac{12 \ep\al \zrd}{\xp} -\frac{18 \al^2\ep^2 \zrz \zrd}{(\xp )^2}
\nn\\
& &+\frac{9\al^4 \ep^4 \zrz \zrd ^2}{(\xp )^3} +...\nn
\eeq
The above asymptotic expansions are found to be a good approximation close to,
but lower in temperature, the maximum of the specific heat.
Once more we are able to present a numerical as well as an analytical
calculation of the relevant quantities.

Suppose that we define $f$ to be the fraction of particles in the ground state as we did in Sec.~4. $N_{ground}$ is given by
\beq
N_{ground}=\Big(e^{\alpha\epsilon}-1\Big)^{-1}\;.\nn
\eeq
From (\ref{h10}), noting that the term in $1/(\epsilon\alpha)$ arises from the ground state, we find the number of particles in excited states is given by
\beq
N_{ex}=\frac{\zeta_R(3)\Omega^3}{\omega_1\omega_2\omega_3}\frac{1}{\alpha^3}+\frac{\zeta_R(2)\Omega^2}{3}\Big(\frac{1}{\omega_1\omega_2}+
\frac{1}{\omega_1\omega_3}+\frac{1}{\omega_2\omega_3}\Big)\left(\frac{3}{2}-\epsilon\right)\frac{1}{\alpha^2}\;.\label{Nex}
\eeq

We will illustrate the results for the case of $N=2000$ as for the isotropic harmonic oscillator using the frequencies for the rubidium experiment. (The results shown are independent of whether the strong or weak trap \cite{BaymPethick} is used, since the difference is one of an overall scaling of the oscillator frequencies, and the results we use are independent of such a scaling.) Fig.~5 shows the result of a comparison of our approximate result for $\epsilon$ (illustrated with diamonds) and the exact result illustrated by the solid curve. Again the agreement is quite good even for relatively large values of $\epsilon$. Fig.~6 shows the comparison between our approximation for the specific heat in (\ref{h12}) and the exact value found from the harmonic oscillator sums. The maximum occurs for $\alpha\simeq0.106$. Fig.~7 shows the ratio of our approximation to the exact specific heat, and for comparison the result of using the bulk expression. As for the isotropic oscillator calculations, the bulk expression shows a significant deviation from the true result; when $\alpha\simeq0.2$ the bulk result is off by about 15\%, whereas our approximation is within about 1\% of the true value. Fig.~8 shows the ratio of the approximate particle numbers to the true value. Our result is seen to have better agreement close to the specific heat maximum, but both our result and the bulk result rapidly converge towards the true value as $\alpha$ increases.

We can now see that the anisotropy has only a small effect on the critical temperature. Using the frequencies $\omega_1=\omega_2=240\pi/\surd8\ {\rm s}^{-1}$, and $\omega_3=240\pi\ {\rm s}^{-1}$, with $\alpha\simeq0.106$ we find $T_m\simeq3.09\times10^{-8}$ K as the temperature at which the specific heat maximum occurs. This can be compared with the temperature of $3.13\times10^{-8}$ K found in the isotropic case in Sec.~4. The bulk temperature (eq.~(\ref{new4.6}) still holds in the anisotropic case with $\omega$ the geometric mean of the frequencies) is about $3.41\times10^{-8}$ K.

\s{Conclusions}

In conclusion,
in this article we presented a detailed analysis of several thermodynamical
quantities for a system of non-interacting spin-0 particles in a general
harmonic oscillator confining potential trap. 
Although there is no phase transition such as that occurring in the free
unconfined boson gas, it is possible to identify a temperature at which BEC
occurs by looking at the maximum in the specific heat. We have seen that this
temperature is nearly identical to the temperature where the groundstate 
occupation starts to increase considerably, the effect actually seen in the
recent experiments through the peak in the velocity distribution of the
sample. 

Attempting to compare the results obtained here directly with the experiments must be done with a certain degree of caution. In the first place we have ignored interatomic interactions, so that there is no distinction made between gases with a positive scattering length \cite{rub,sod}, and those with a negative scattering length \cite{lith}. Secondly, it is perhaps not quite so clear that the use of the grand canonical ensemble is justified for systems with such a relatively small number of particles \cite{GHmicro}. With these caveats in mind, for the case of rubidium we found the specific heat maximum to occur at $T\simeq31$ nK for 2000 particles. For the case of sodium, if we use $N=2.5\times10^5$ we find the specific heat maximum to occur at $T\simeq1.16\ \mu{\rm K}$. The results for the temperature found from using the bulk approximation were very close to these values, so it is unlikely that the present experiments can distinguish between the bulk approximation, our approximation or the exact value. It was apparent from our calculations that the specific heat found from our approximation was much closer to the exact result than the bulk approximation was. Perhaps in future experiments it will be possible to provide a more stringent test of the various approximations.

We now wish to mention the comparison between our results and the density of states method used in Ref.~\cite{PhysLett,GH}. In this approach the authors separated off the ground state contribution for the particle number and treated the remaining terms in the sum by approximating it with an integral over the energy. The density of states was parametrized by
\beq
\rho(E)=\frac{1}{2}\frac{E^2}{(\hbar\omega)^3}+
\gamma\frac{E}{(\hbar\omega)^2}\;,\label{6.1}
\eeq
where $\omega=(\omega_1\omega_2\omega_3)^{1/3}$ and $\gamma$ is a dimensionless function of the frequencies. In the isotropic case $\gamma=3/2$, but the authors \cite{PhysLett,GH} had to determine $\gamma$ numerically in the anisotropic case. The bulk approximation we mentioned earlier consists of ignoring the term in $\gamma$. By contrast, the approximate method we presented is entirely analytical with no numerical evaluations required. By comparing the results of our calculation for the particle number with those of Ref.~\cite{PhysLett,GH} we can deduce an analytic value for $\gamma$. We find
\beq
\gamma=\frac{1}{2}\omega^2\left(\frac{1}{\omega_1\omega_2}+
\frac{1}{\omega_2\omega_3}+\frac{1}{\omega_3\omega_1}\right)\;.\label{6.2}
\eeq
This value has also been borne out by an independent evaluation \cite{KKDJTstates} of the density of states which in addition obtains the next order correction to (\ref{6.1}) in the anisotropic case.

Although the gases in the experiments are dilute, 
in order to get a quantitatively more satisfactory picture for the recent
experiments, the vapor has to be treated as a weakly interacting system. In 
the quantum field theory approach to BEC this might be done in a systematic
way \cite{KKDJTreview}. One possibility is to treat the interaction as a 
perturbation and calculate the leading corrections to the free boson gas 
treated in the present article. The detailed knowledge of the lowest order
in perturbation theory provided here is thus an important basis for future
developments in this direction. 
Another possibility is to consider an effective theory for the 
groundstate, its occupation number being a very good indicator for the onset
of BEC.

After this paper was submitted for publication, another independent calculation of BEC in harmonic oscillator confining potentials appeared \cite{HHR}. This paper uses the Euler-Maclaurin summation formula to evaluate the harmonic oscillator sums. The result is expressed in terms of polylogarithms, much like the results found in our Appendix~D. One difference between this approach and ours is that the authors of Ref.~\cite{HHR} introduce an effective fugacity which means that the argument of their polylogarithms never becomes equal to unity. It is straightforward to modify the approach we have discussed in Appendix~D and obtain in an easy way results which appear to be equivalent to those of Ref.~\cite{HHR}. However this method necessarily involves a numerical evaluation of the polylogarithms, as we mentioned earlier in connection with our own polylogarithm results, and is therefore not entirely analytic.

\ack{We are indebted to Stuart Dowker, with whom we learnt many properties
of the Barnes zeta functions. Furthermore we thank Michael Bordag, 
Volkhard M\"uller and Wolfgang Weller for interesting discussions. We are also grateful to L.~D.~L.~Brown for advice about the numerical procedures used. This investigation has been supported by the DFG under the contract number Bo 1112/4-1.}

\appendix
\app{The Barnes zeta function}
As we have seen, in order to determine the asymptotic expansion of some
thermodynamical quantities, we need several properties of the Barnes
zeta function \cite{barnes03,dowker94},
\beq
\zb s,\ar =\sum_{\vec m =0}^{\infty} (a+\vec m \vec r)^{-s} ,\label{a1}
\eeq
with $\vec r$ a d-dimensional vector. 
In equation (\ref{eq22}) we used the notation 
$\zb s,a,d) $ for $\vec r =\vec 1$. The residues of $\zb s, \ar$ at 
$s=1,...,d$ and the values of the function at $s=-p$, $p\in \nats_0$ are most
easily deduced using the representation as a contour integral
\beq
\zb s, \ar =\frac{i\Gamma (1-s)}{2\pi} \int_C dt\,\,
(-t)^{s-1} \frac{e^{-at}}{\prod_{i=1}^d \left(
1-e^{-r_i t}\right)},\label{a2}
\eeq
where the contour $C$ is counterclockwise enclosing the positive real
axis. The only possible pole occurs at $t=0$. For that reason one might
like to introduce the generalized Bernoulli polynomials \cite{nor} through
\beq
 \frac{e^{-at}}{\prod_{i=1}^d \left(
1-e^{-r_i t}\right)} = \frac{(-1)^d}{\pri} 
\sum_{n=0}^{\infty} \ber _n (\ar \frac{ (-t)^{n-d}}
{n!}.\label{a3}
\eeq
In terms of these it is immediate that for $n=1,...,d$,
\beq
\rzb n, \ar =\frac{(-1)^{d+n}}{(n-1)!(d-n)!\pri} \ber _{d-n} (\ar \label{a4}
\eeq
and for $p\in\nats_0$,
\beq
\zb -p, \ar =\frac{(-1)^dp!\ber _{d+p} (\ar }{\pri (d+p)!}.\label{a5}
\eeq
For the leading asymptotics of the thermodynamical quantities we need at 
most the first three residues in (\ref{a4}). These are given explicitly
by
\beq
\rzb d,\ar &=&\frac 1 {(d-1)!\pri},\nn\\
\rzb d-1, \ar &=& \frac{\sri -2a}{2(d-2)!\pri}, \label{a6}\\
\rzb d-2,\ar &=& \frac{6a^2 -6a\sri +\left( \sri \right)^2 +
\sum_{i,j=1,i<j}^d r_ir_j}{12(d-3)!\pri}.\nn
\eeq
In addition to the above equation we needed only 
\beq
\zb s,\ar =\frac 1 {a^s} +\cao (a^0),\label{a7}
\eeq
which is obvious from the original sum (\ref{a1}). Using equations (\ref{a6}) and (\ref{a7}) we found the
asymptotic expansion for all thermodynamical quantities.
\app{Asymptotics for $x\ll 1$, $\ep \ll 1$ 
of the sums $u_1$, $S_1$, $S_2$ and $S_3$}
In this appendix we list the results used for the derivation of the internal
energy and the specific heat of the isotropic harmonic oscillator confining
potential. We needed the following asymptotic expansions~:
\beq
u_1&=&f(0,4,1)\nn\\
&=&\frac{\zrv}{x^4} +\frac{(1-\ep) \zrd}{x^3} +\frac
1 3 \frac{\zrz}{x^2}+\cao \left(\frac {\ep}{ x^2}, \frac 1 x\right)\;,\nn\\
S_1&=&f(-1,3,0)\nn\\
&=&\frac 1 {x^2\ep^2}+\frac{\zrz}{x^3}+\cao \left(\frac{\ln x}{x^2}
\right)\;,\nn\\
S_2&=&f(-1,4,1)\nn\\
&=&\frac{\zrd}{x^4}+\frac{\zrz}{x^3}+\cao\left(
\frac{\ln x}{x^2},\frac{\ep}{ x^3}\right)\;,\nn\\
S_3&=&f(-1,5,2)\nn\\
&=&\frac{\zrv}{x^5} +\frac{\zrd}{2x^4}-\frac{\ep \zrd}{x^4} -\frac 1 {12} 
\frac{\zrz}{x^3}+\cao \left(\frac{\ln x}{x^2},\frac{\ep}{x^3}\right)\;.\nn
\eeq
These results lead, after some calculation, to equations (\ref{eq28}) and (\ref{eq29}).

\app{Asymptotics for $x\ll 1$, $\ep \ll 1$ 
of the sums $u_i$, $S_1$, $S_{2,i}$ and $S_{3,ij}$}
In this appendix we give the results used for the derivation of the
asymptotics of several thermodynamical quantities. (See equations 
(\ref{h9})-(\ref{h12}).)

First of all we need the analogous results to equations (\ref{eq24}) and
(\ref{eq25}) for the anisotropic oscillator. Unfortunately for the 
anisotropic oscillator it is not possible to write all needed sums in a
unified form as done in equation (\ref{eq18}). For that reason we
have to list several results. The techniques are exactly the same 
techniques as those employed in section 3. We found for $l=-1,0$
\beq
\sneu \frac{\en}{n^l \pxj}&=& \sum_{m=1 \atop m\neq 1-l}^3 
\Gamma (m) \zeta_R (m+l) \rzb m,\ep\al|\vec x)\nn\\
& &+\Gamma (1-l) \left\{\mbox{PP }\zb 1-l,\ep\al|\vec x)\right.\nn\\
& &\left.\qquad +\left(\gamma +\psi (1-l)\right) 
\rzb 1-l,\ep\al |\vec x )\right\}
\nn\\
& &+\zeta_R(l) \zb 0,\ep\al|\vec x )+...,\nn
\eeq
whereas for $l=1$ one has
\beq
\sneu \frac{\en}{n\pxj} &=&\sum_{m=1}^3 \Gamma (m) \zeta_R (1+m) \rzb m,
\ep\al|\vec x ) \nn\\
& &+\zeta_{\cab} ' (0,\ep\al|\vec x )+...\;.\nn
\eeq
These results can be used for the calculation of $q,N$ and $S_1$.

For $u_i$ and $S_2,i$ one needs
\beq
\lefteqn{\sneu \frac{\enx}{\pxj}\frac 1 {\exi}=}\hspace{1cm}\\
&&\sum_{m=1\atop m\neq 1-l}^4\Gamma (m) \zeta_R (m+l) \rzb (m,\ep\al +x_i|
(\vec x ,x_i))\nn\\
&&+\Gamma (1-l) \Big\{ \mbox{PP } \zb 1-l,\ep\al+x_i|(\vec x ,x_i))
 \Big(\gamma +\psi (1-l) \Big)\nn\\
&&\times\rzb 1-l,\ep\al +x_i|
(\vec x ,x_i ))\Big\}+...\;.\nn
\eeq
Finally for $S_{3,ij}$ we need
\beq
\lefteqn{
\sneu \frac{ne^{-n\al\ep -n(x_i+x_j)}}{\prod_{l=1}^3 \left(1-e^{-nx_l}
\right)}\frac 1 {\left(1-e^{-nx_i}\right)}\frac 1 
{\left(1-e^{-nx_j}\right)}   }\nn\\
& &=\sum_{m=1\atop m\neq 2}^5 \Gamma (m) \zeta_R (m-1) \rzb m,\ep\al+x_i
+x_j | (\vec x ,x_i,x_j ))\nn\\
& &+\mbox{PP } \zb 2,\ep\al+x_i+x_j | (\vec x ,x_i,x_j))\nn\\
& &+\rzb 2,\ep\al+x_i+x_j|( \vec x ,x_i,x_j))+...\;.\nn
\eeq
These results are enough to find the following expansions,
\beq
u_i &=& \frac{\zrv}{ \xp \al} +\frac{\zrd}{2\xp } \left(3 -\frac {x_i}
{\al} -2\ep\right) \nn\\
& &+\frac{\zrz \al}{12 \xp} \left( 9-9\frac{x_i}{\al} +\frac{\xij +x_i^2}
{\al^2} \right),\nn\\ 
S_1 &=& \frac 1 {(\al\ep )^2} +\frac{\zrz }{\xp}+...,
\nn\\
S_{2,i}&=& \frac{\zrd }{\xp x_i} +\frac{\zrz \al}{ 2\xp x_i} 
\left( 3 -\frac {x_i}{\al } \right) +...,\nn\\
S_{3,ij} &=& \frac{\zrv}{ \xp x_i x_j } +\frac{ \zrd \al } {\xp x_i x_j }
\left( \frac 3 2 -\frac {x_i+x_j}{2\al} -\ep \right) 
+\frac{ \zrz \al^2} {12 \xp x_i x_j } \nn\\
&&\quad\times\left( 9 -9\frac{x_i+x_j} {\al}
 +\frac{\xij +x_i^2 +x_j^2 +3x_i x_j }{\al^2 } \right) +...\;.\nn
\eeq
\app{Asymptotics for $x\ll 1$ for the anisotropic harmonic oscillator}
As mentioned in section 3 and 4, it is possible to obtain an asymptotic
expansion valid for $x\ll 1$ without restricting $\ep$ to the range of
small parameters. The way how to obtain the approximation is described at
the end of section 3. The results analogous to equation (\ref{eq24}) and  
(\ref{eq25}) read
\beq
\sneu \frac{n^{-l} \en}{\pxi} &=&
\sied \Gamma (i) \rzb i,0|\vec x ) Li_{l+i} \liar +...\;,\nn
\eeq
\beq
\sneu \frac{n^{-l} \enx}{\exi \pxj }&=& 
\sum_{n=1}^4 \Gamma (n) \rzb n,x_i|(\vec x,x_i )) Li_{l+n}\liar+...\;,\nn
\eeq 
\beq 
\lefteqn{\sneu \frac{n e^{-n\al\ep -n (x_i+x_j) }}{\exi \exj \prod_{l=1}^3
\left(1-e^{-nx_l}\right)}=} \\
&&\sum_{n=1}^5 \Gamma (n)\times
 \rzb n,x_i+x_j|(\vec x,x_i,x_j)) Li_{n-1}\liar 
+...\;.\nn 
\eeq
As a result, the following asymptotics for the thermodynamical quantities
are derived,
\beq
q&=&\frac 1 {\xp} Li_4 \liar +\frac 3 2 \frac{\al}{\xp} Li_3 \liar 
\nn\\
& &+\left( \frac 3 4 \frac{\al^2}{\xp} +\frac 1 {12} \frac{\xij}{\xp}
\right) Li_2 \liar +...\;,\nn\\
N&=& \frac 1 {\xp} Li_3 \liar +\frac 3 2 \frac{\al}{\xp} Li_2 \liar\nn
\\
& &+\left(\frac 3 4 \frac{\al^2}{\xp} +\frac 1 {12} \frac{\xij}{\xp}
\right) Li_1 \liar +...\;,\nn\\
\frac U {\hbar \Om } &=& \frac 3 {\al \xp} Li_4 \liar +\frac 9 {2 \xp }
Li_3 \liar \nn\\
& &+\frac 1 {12\al \xp} (36\al^2 +\xij ) Li_2 \liar +...\;,\nn\\
\frac C k &=& \frac 1 {\xp} \left\{
-\frac {9Li_3^2 \liar }{Li_2 \liar} +12 Li_4 \liar \right\}\nn\\
& &+\frac{\al}{\xp} \left\{
-9Li_3 \liar +\frac{27} 2 \frac{Li_1 \liar Li_3^2\liar}{Li_2^2 \liar}\right\}
+...\;.\nn
\eeq
This concludes the list of asymptotic expansions which we are going to 
give in the present article.

\eject
\begin{center}
{\bf\large Figure Captions}
\end{center}
\bigskip
\begin{description}
\item [Figure 1] This shows $\epsilon$ as a function of $x=\hbar\omega/(kT)$. The solid curve shows the result found from a numerical evaluation of the exact harmonic oscillator sums for the isoptropic case for $N=2000$. The diamonds show the result of using our analytic approximation.
\item [Figure 2] The specific heat computed numerically for the isotropic harmonic oscillator is shown as the solid curve. The diamonds show the result using our approximation. The units for the specific heat are in factors of the Boltzmann constant $k$. The particle number is $N=2000$. The maximum occurs for $x\simeq0.0921$. Our approximation breaks down for $x$ below the point where the specific heat maximum occurs.
\item [Figure 3] The diamonds show the ratio of our approximation for the specific heat to the exact result. The crosses denote the ratio of the bulk specific heat to the exact value. $N=2000$ is taken. Both results become increasingly inaccurate below the specific heat maximum.
\item [Figure 4] The ratio of the approximate to the exact ground state particle number is shown. The diamonds illustrate the result of using our approximation and the crosses denote the results found using the bulk approximation.
\item [Figure 5] The numerical result for $\epsilon$ is plotted against $\alpha=\hbar(\omega_1+\omega_2+\omega_3)/(3kT)$ for the anisotropic oscillator for $N=2000$. The frequencies are taken to be $\omega_1/(2\pi)=\omega_2/(2\pi)=42.4$ Hz and $\omega_3/(2\pi)=120$ Hz. The diamonds show the result found from using our approximation.
\item [Figure 6] The exact value for the specific heat is shown as the solid curve, and our approximation as the diamonds. $N=2000$ and the frequencies are as in the previous figure.
\item [Figure 7] The ratio of our approximate specific heat to the exact value is shown with diamonds. The result found from using the bulk result is shown with crosses.
\item [Figure 8] The ratio of the ground state occupation number found using approximation to the exact result is shown by the diamonds. The ratio of the bulk specific heat to the exact value is shown by the crosses. The frequencies and particle numbers are the same as the previous three figures.
\end{description}
\eject
\begin{figure}[ht]
\begin{center}
\leavevmode
\epsffile{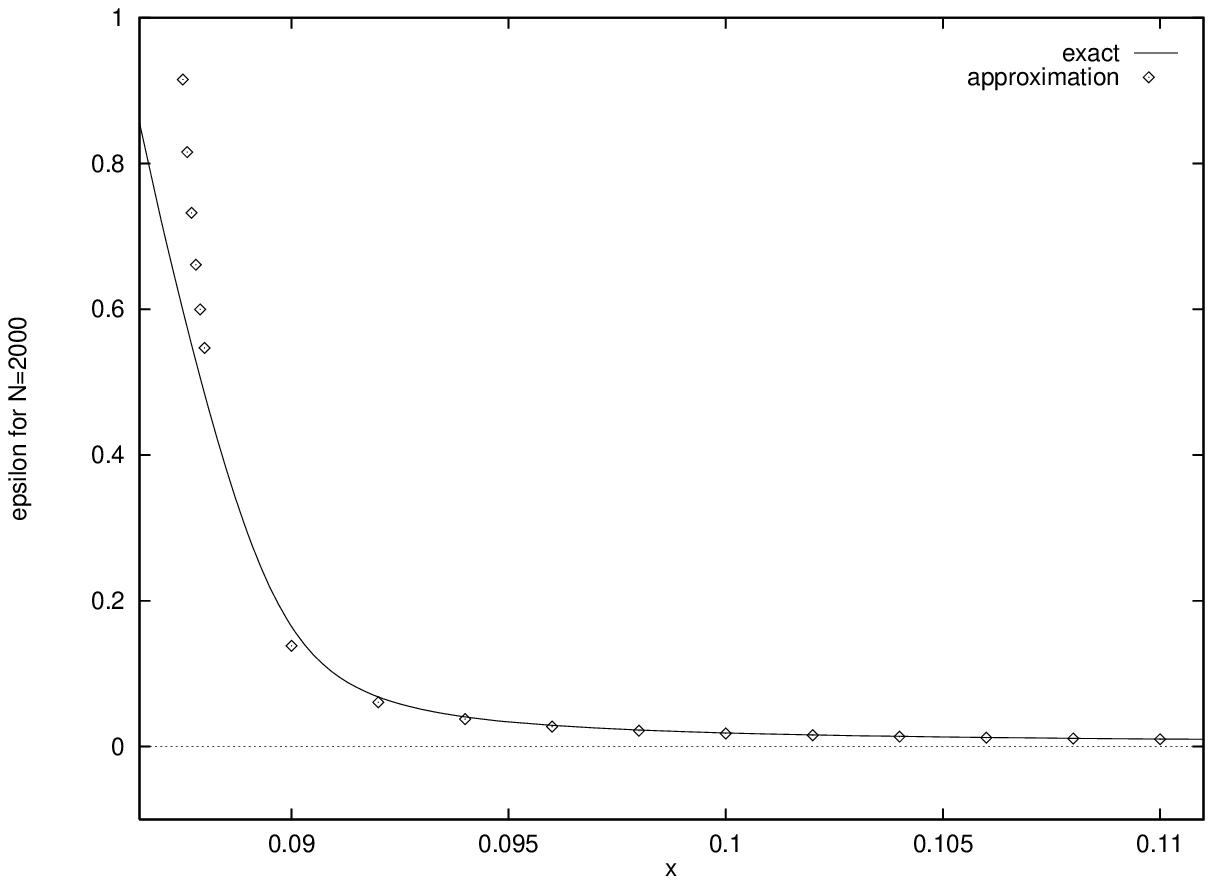}
\end{center}
\caption{}
\end{figure}
\eject
\begin{figure}[ht]
\begin{center}
\leavevmode
\epsffile{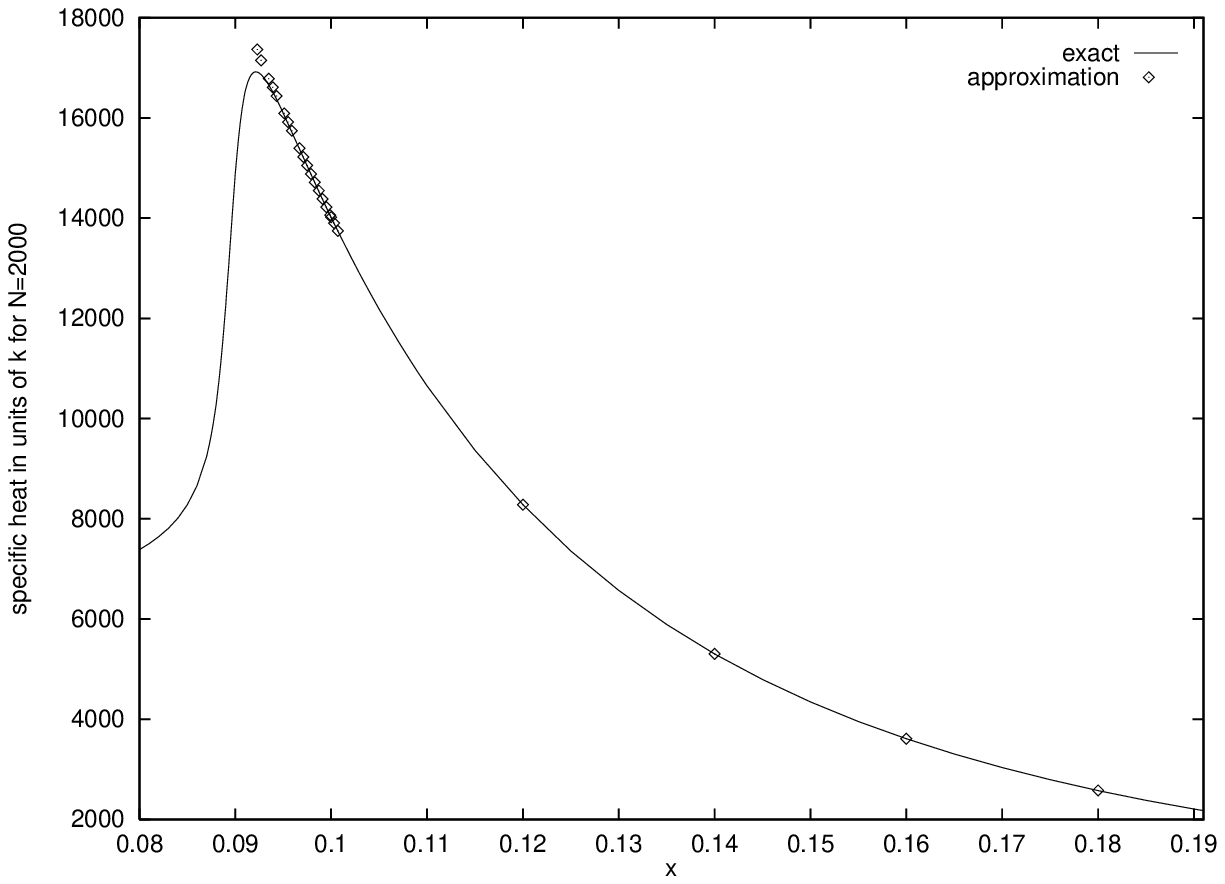}
\end{center}
\caption{}
\end{figure}
\eject
\begin{figure}[ht]
\begin{center}
\leavevmode
\epsffile{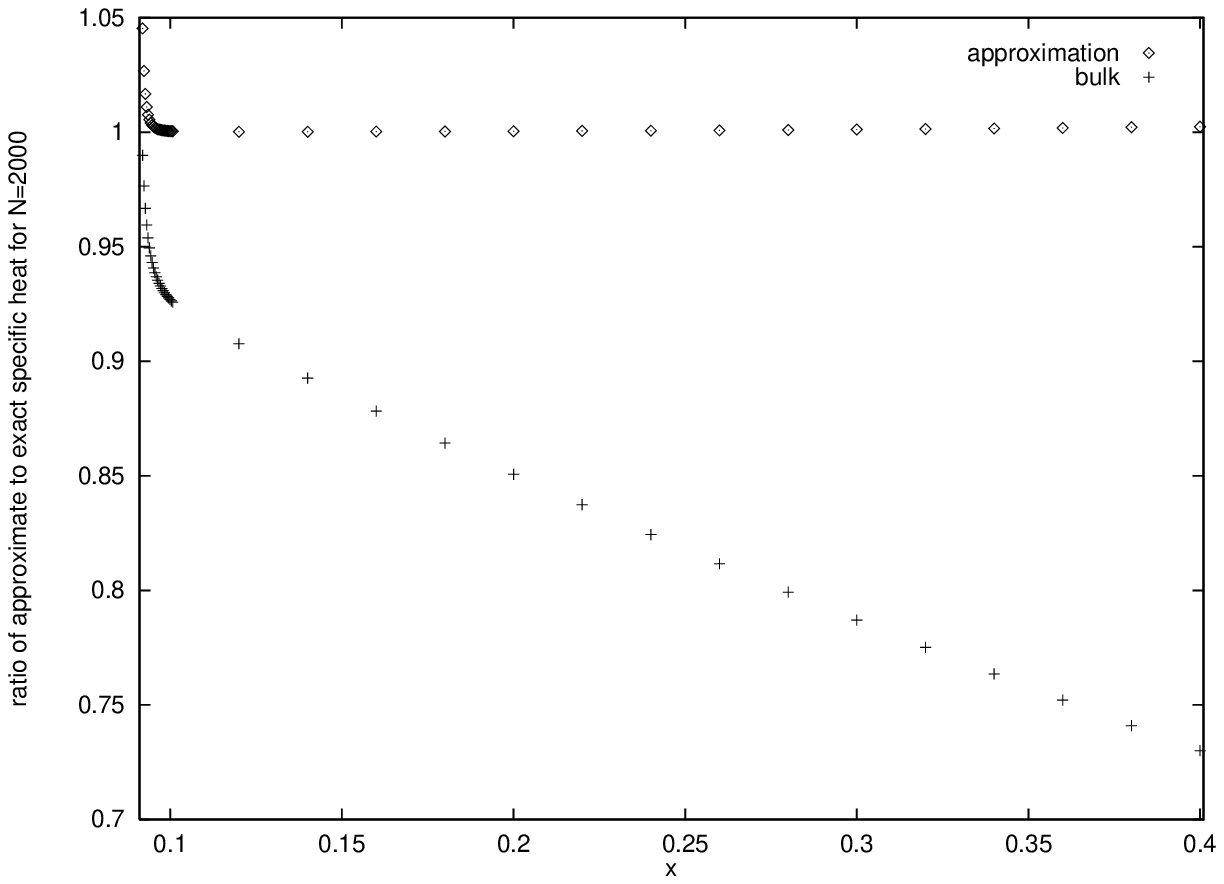}
\end{center}
\caption{}
\end{figure}
\eject
\begin{figure}[ht]
\begin{center}
\leavevmode
\epsffile{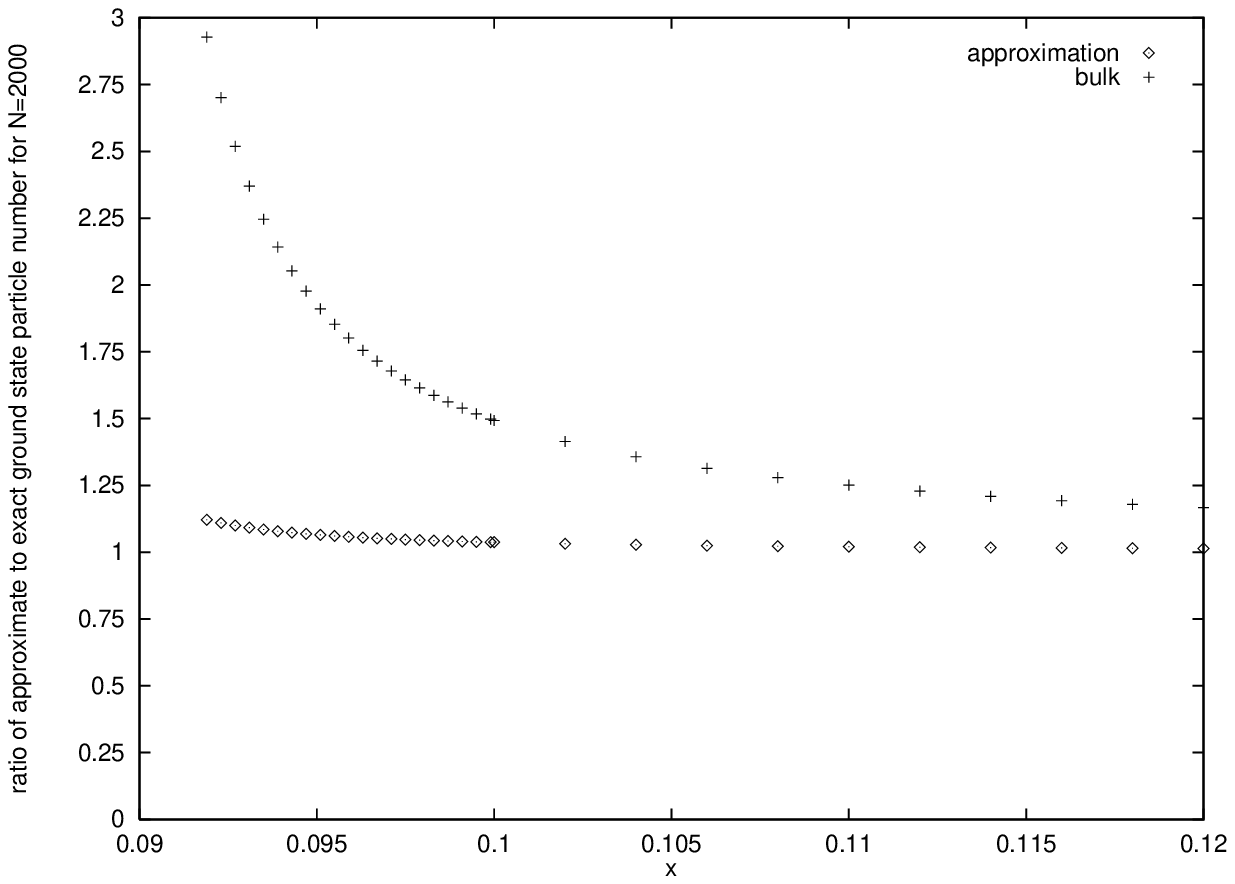}
\end{center}
\caption{}
\end{figure}
\eject
\begin{figure}[ht]
\begin{center}
\leavevmode
\epsffile{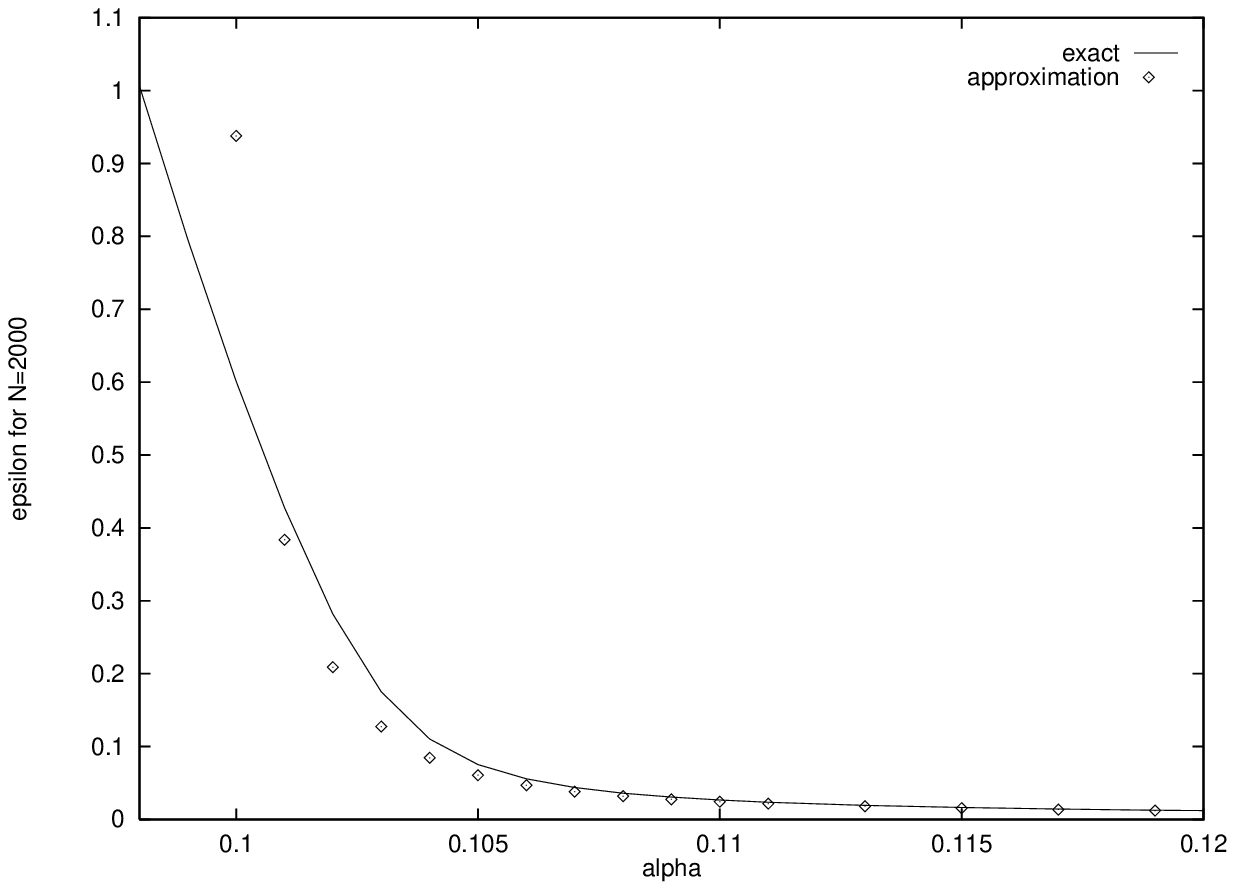}
\end{center}
\caption{}
\end{figure}
\eject
\begin{figure}[ht]
\begin{center}
\leavevmode
\epsffile{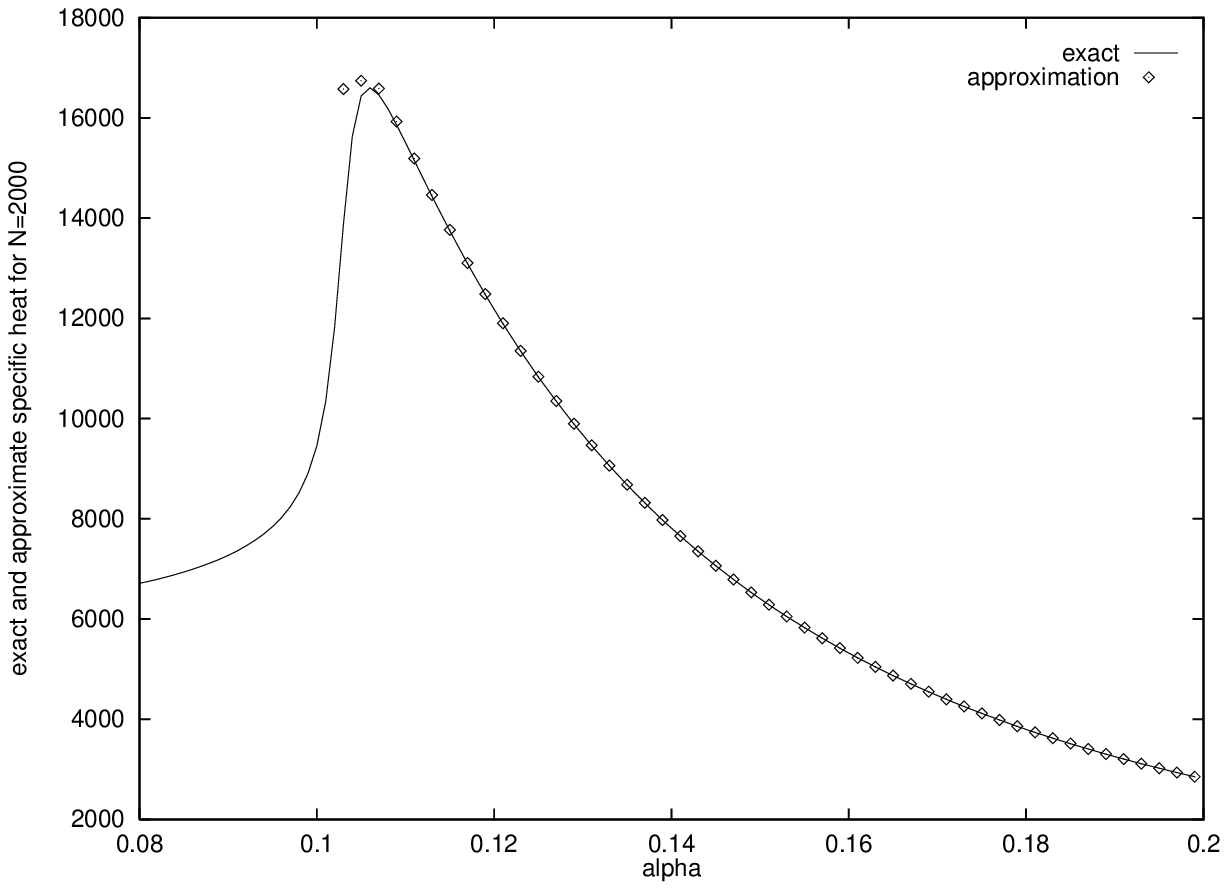}
\end{center}
\caption{}
\end{figure}
\eject
\begin{figure}[ht]
\begin{center}
\leavevmode
\epsffile{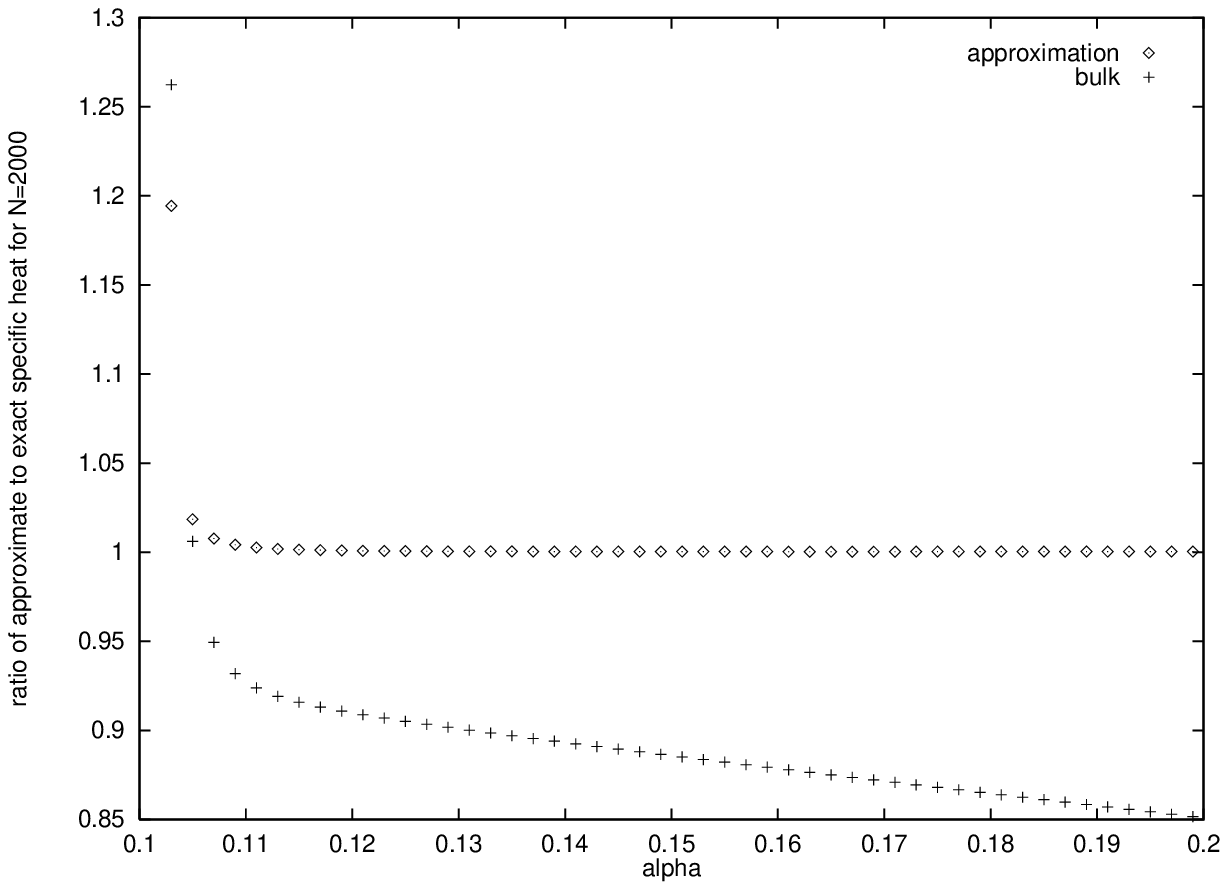}
\end{center}
\caption{}
\end{figure}
\eject
\begin{figure}[ht]
\begin{center}
\leavevmode
\epsffile{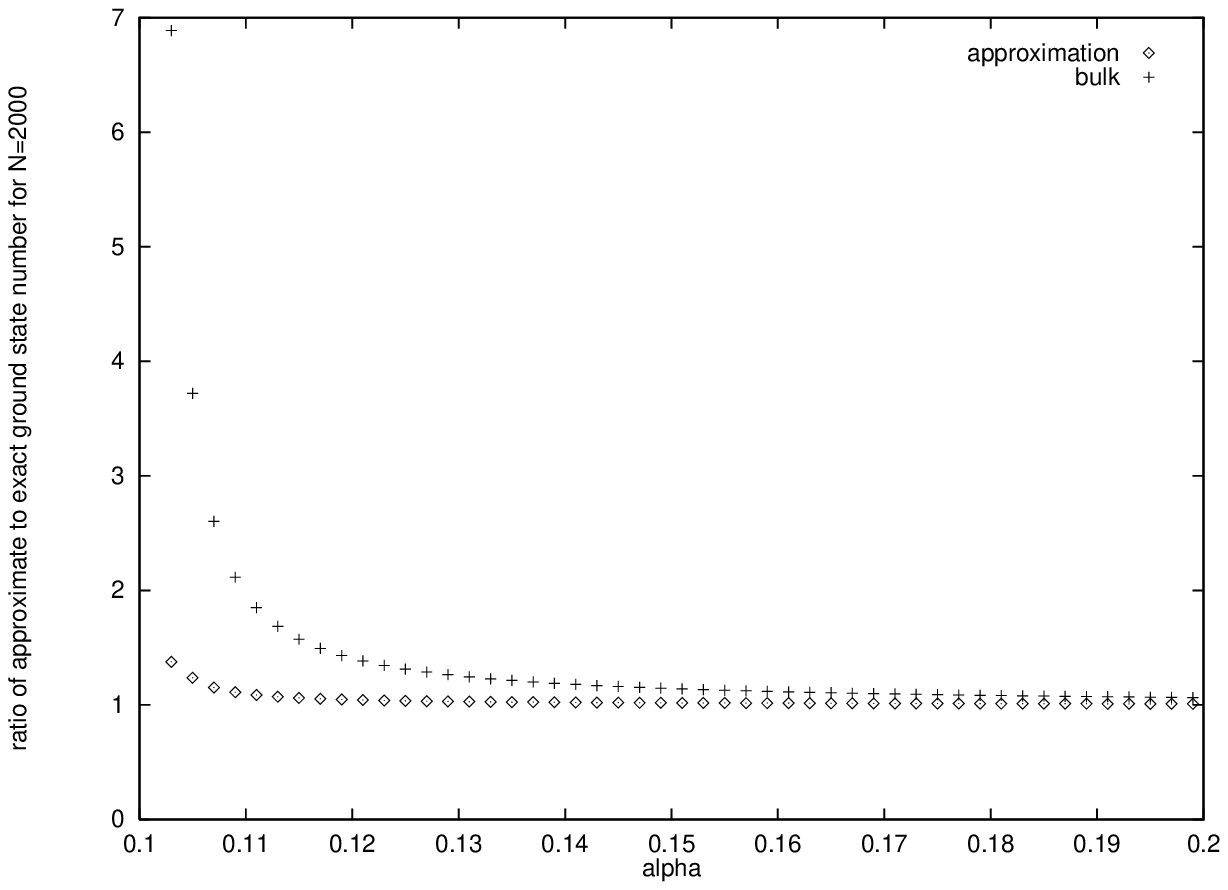}
\end{center}
\caption{}
\end{figure}
\eject
\end{document}